\documentclass[10pt]{JHEP3}

\usepackage{epsfig,multicol,bbm}
\usepackage{amsmath}
\usepackage{graphicx}
\usepackage{dcolumn}
\usepackage{bm}
\usepackage{amssymb}
\usepackage{latexsym}

\newcommand\fverb{\setbox\fverbbox=\hbox\bgroup\verb}
\newcommand\fverbdo{\egroup\medskip\noindent%
            \fbox{\unhbox\fverbbox}\ }
\newcommand\fverbit{\egroup\item[\fbox{\unhbox\fverbbox}]}
\newbox\fverbbox

\title{Chaplygin gas and the cosmological evolution of alpha}

\author{Chao-Jun Feng\\
    Shanghai United Center for Astrophysics(SUCA), Shanghai Normal University,\\
    100 Guilin Road, Shanghai 200234, P.R.China\\
    E-mail: \email{fengcj@shnu.edu.cn}}

\author{Xin-Zhou Li\\
    Shanghai United Center for Astrophysics(SUCA), Shanghai Normal University,\\
    100 Guilin Road, Shanghai 200234, P.R.China\\
    E-mail: \email{kychz@shnu.edu.cn}}

\received{\today}       
\accepted{\today}       

\abstract{The class of Chaplygin gas models regarded as a candidate of dark energy can be realized by a scalar field,
which could drive the variation of the fine structure constant $\alpha$ during the cosmic time. This phenomenon has
been observed for almost ten years ago from the quasar absorption spectra and attracted many attentions. In this paper,
we reconstruct the class of Chaplygin gas models to a kind of scalar fields and confront the resulting
$\Delta\alpha/\alpha$ with the observational constraints. We found that if the present observational value of the
equation of state of the dark energy was not exactly equal to $-1$, various parameters of the class of Chaplygin gas
models are allowed to satisfy the observational constraints, as well as the equivalence principle is also respected. }


\dedicated{Dedicated to the people's republic of China \\on her 60th anniversary 1949-2009.}

\begin{document}


\newpage
\section{Introduction}

The possibility of varying the fundamental constants over cosmological time-scale has been studied for many years
\cite{Uzan}. Among these fundamental constants, the time-variation of the fine-structure constant $\alpha$ is deserved
to study both from the experimental and theoretical point of view. The observational evidence of time-varying $\alpha$
is firstly from the quasar absorption spectra and reported in 2001 by Webb et al., and now there are  some other
independently results can be used to constrain the variation of $\alpha$, see \cite{Uzan}, \cite{Olive} and
\cite{Martins}. The variation of $\alpha$ can be due to many possible reasons, one of them is that there is a scalar
field coupled to the gauge field. In 1982, Bekenstein first introduced the exponential form for the coupling, which in
practice can be taken in the linear form coupling between a scalar field and the electromagnetic field to explain the
variation.

As we mentioned above, there are several observational constraints on the variation of $\alpha$ as the function of
redshift $z$, namely
\begin{equation}
    \frac{\Delta\alpha(z)}{\alpha} \equiv \frac{\alpha(z)-\alpha_0}{\alpha_0} \,,
\end{equation}
where $\alpha_0 = \alpha(0)$ denotes its present measured value. After Webb et al.'s report, the Oklo natural fission
reactor found the variation of $\alpha$ with the level $|\Delta\alpha/\alpha| \lesssim 10^{-7}$ at the redshift
$z=0.14$ \cite{Olive}, \cite{Damour}. The computation of ${}^{187}$Re half-life in meteorites gives
$|\Delta\alpha/\alpha| \lesssim 10^{-7}$ at the redshift $z=0.45$ \cite{Olive}. The absorption line spectra of distance
quasars suggests $\Delta\alpha/\alpha = (-0.543\pm0.116)\times10^{-5}$ between $z=0.2$ and $z=3.7$ \cite{Webb}. And the
recent detailed analysis of high quality quasar spectra gives the lower variation $\Delta\alpha/\alpha
=(-0.06\pm0.06)\times 10^{-5}$ over the redshift $z=0.4-2.3$ and $\Delta\alpha/\alpha =(0.15\pm0.43)\times 10^{-5}$
over the redshift $z=1.59-2.92$ \cite{Chand}. In the following, we only consider the upper limit for the variation of
$\alpha$, so we have chosen the conservative constraint $|\Delta\alpha/\alpha| \lesssim 10^{-6}$ over the redshift $z =
0.4-3$. The limit from the power spectrum of anisotropies in the Cosmic Microwave Background (CMB) is
$|\Delta\alpha/\alpha| \lesssim 10^{-2}$ at $z= 10^3$ \cite{Avelino}. At last, the most ancient data from Big Bang
Nucleosynthesis is $|\Delta\alpha/\alpha| \lesssim 10^{-2}$ over the redshift $z=10^8-10^{10}$ \cite{Avelino},
\cite{Nollett:2002da}.

Quite independently, observations like Type Ia supernovae, CMB and SDSS et al. have strongly confirmed that our
universe is accelerated expanding recently caused by an unknown energy component called dark energy. Experiments have
indicated there are mainly about $73\%$ dark energy and $27\%$ matter components in the recent universe, but so far
people still do not understand what is dark energy from fundamental theory. The best candidate seems the cosmological
constant including the vacuum energy, but it suffers the fine-tuning and coincidence problems. In order to alleviate
these problems, a kind of scalar models called quintessence is needed to explain the origin of the dark energy. Thus it
is natural to consider that quintessence or other type of scalar field models could be responsible for the time
variation of $\alpha$ and it is a possible way to to distinguish dynamical dark energy models from the a cosmological
constant. For the recent progress on the varying alpha, see \cite{Wei:2009xg} and references therein.

Another candidate class of dark energy model is called the Chaplygin gas model \cite{Kamenshchik:2001cp}, which has
been developed for many years, see \cite{Bento:2002ps}, \cite{Bento:2002uh} and \cite{Benaoum:2002zs}. Since Chaplygin
gas can be always described by a scalar field with a effective potential, which is a process called reconstructing a
scalar field from Chaplygin gas. So, it is natural to consider the time variation of $\alpha$ driven by Chaplygin gas
which realized by a scalar field. In this paper, we first briefly review how a scalar field (quintessence) to drive a
time-varying $\alpha$ in the next section. In Section 3. various Chaplygin gas models are reviewed and we also
reconstruct them to scalar fields. In Section 4. we show the variation of $\alpha$ driven by Chaplygin gas and the
conclusion is given in the last section.

\section{Varying alpha from quintessence}
Let us consider the following action
\begin{equation}\label{action}
    \mathcal{S} = \frac{1}{2} \int d^4x \sqrt{-g}~\mathcal{R}
    + \int d^4x\sqrt{-g}~\mathcal{L}_{\phi}(\phi)
    + \int d^4x\sqrt{-g}~\mathcal{L}_{F}(\phi, F_{\mu\nu}) + \mathcal{S}_m \,,
\end{equation}
where $\mathcal{L}_{\phi}$ is the Lagrangian density for the quintessence field $\phi$ minimally coupled to gravity as
the following
\begin{equation}\label{Lag scalar}
    \mathcal{L}_{\phi} = \frac{1}{2}\partial^\mu\phi\partial_\mu\phi - V(\phi) \,,
\end{equation}
and $\mathcal{S}_m$ is the action for ordinary pressureless matter. Here, $\mathcal{L}_{F}$ is the Lagrangian density
for an electromagnetic field $F_{\mu\nu}$ coupled to quintessence field as
\begin{equation}\label{Lag mag}
    \mathcal{L}_F(\phi) = -\frac{1}{4}B_F(\phi)F_{\mu\nu}F^{\mu\nu} \,,
\end{equation}
where $B_F(\phi)$ allows for the evolution in $\phi$ and $B_F(\phi_0) = 1$, where the subscript $0$ represents the
present value of the quantity. The effective fine structure constant depends on the value of $\phi$ as
\begin{equation}\label{fine stru}
    \alpha = \frac{\alpha_0}{B_F(\phi)} \,,
\end{equation}
and thus we have
\begin{equation}\label{delta fine stru}
    \frac{\Delta\alpha}{\alpha} \equiv \frac{\alpha-\alpha_0}{\alpha_0} = \frac{1-B_F(\phi)}{B_F(\phi)} \,.
\end{equation}
We will consider a spatially flat Friedmann-Robertson-Walker(FRW) universe with the metric $ds^2 =
dt^2-a^2(t)d\bf{x}^2$ and assume the scalar field is homogeneous during the evolution of universe, then the energy
density and pressure of the scalar field are
\begin{equation}\label{energy density and pressure}
    \rho_\phi = \frac{\dot\phi^2}{2} + V(\phi) \,, \quad  p_\phi = \frac{\dot\phi^2}{2} - V(\phi) \,,
\end{equation}
where a dot denotes the derivative with respect to the cosmic time $t$. And the equations of motion are
\begin{eqnarray}
  \dot H &=& -\frac{1}{2}\bigg(\rho_m + \dot\phi^2\bigg) \,, \\
  \dot \rho_m &=& - 3H\rho_m \label{matter cons}\,,\\
  \ddot \phi &=& -3H\dot \phi - \frac{dV(\phi)}{d\phi} \label{phi eom}\,,
\end{eqnarray}
subject to the Friedmann constraint
\begin{equation}\label{Fried equ}
    H^2 = \frac{1}{3}\bigg(\rho_m + \rho_\phi\bigg)\,,
\end{equation}
and the solution to eq.(\ref{matter cons}) is simply $\rho_m = \rho_{m0}a^{-3}$. In fact, the equation of motion
(\ref{phi eom}) for $\phi$ should be added a term proportional to $F_{\mu\nu}F^{\mu\nu}$ and the derivative of $B_F$.
However, such a term can be safely neglected due to the following reasons. First, the derivative of $B_F$ actually
corresponds to the time derivative of $\alpha$, which is very small when we consider the equivalence principle
constraints. Second, the statistical average of the term $F_{\mu\nu}F^{\mu\nu}$ over a current state of the universe is
zero.

\section{Chaplygin gas models}

\subsection{Chaplygin gas}
There exist an interesting class of dark energy models involving a fluid known as a Chaplygin gas
\cite{Kamenshchik:2001cp}, which can explain the acceleration of the universe at later times and its equation of state
is
\begin{equation}\label{chaplygin gas}
    p = -\frac{A}{\rho}\,,
\end{equation}
which can be obtained from the Nambu-Goto action for $d$-branes moving in a $(d+2)$-dimensional spacetime in the
light-cone parametrization. With the equation of state (\ref{chaplygin gas}) the energy conservation law $d(\rho a^3) =
-pd(a^3)$ can be integrated to give
\begin{equation}\label{chaplygin gas energy density }
    \rho = \sqrt{A + \frac{B}{a^6}} \,,
\end{equation}
where $B$ is an integration constant. By choosing a positive value for $B$, we can find that $\rho \sim \sqrt{B}/a^3$
when $a$ is small ($a\ll(B/A)^{1/6}$) and $\rho \sim -p\sim\sqrt{A}$ when $a$ is large ($a\gg(B/A)^{1/6}$). Thus, at
earlier times when $a$ is small, the gas behaves like a dust (pressureless) and it behaves as a cosmological constant
at late times, thus leading to an accelerated expansion. In a generic situation, there is an intermediate phase, in
which it looks like a mixture of a cosmological constant with a "stiff" matter ($p=\rho$).

One can obtain a homogeneous scalar field $\phi(t)$ with its potential $V(\phi)$ and Lagrangian density (\ref{Lag
scalar}) to describe the Chaplygin cosmology by setting the energy density and pressure of the field (\ref{energy
density and pressure}) equal to that of the Chaplygin gas and we find
\begin{eqnarray}
  \dot \phi^2 &=& \frac{B}{a^6\sqrt{A+B/a^6}} \label{phi rec 1}\,,\\
  V(\phi) &=& \frac{1}{2}\bigg(\sqrt{A+B/a^6} + \frac{A}{\sqrt{A+ B/a^6}}\bigg)\,.
\end{eqnarray}
By using the Friedmann equation (\ref{Fried equ}), we  get the variation of $\phi$ in terms of the integration of $a$
from eq.(\ref{phi rec 1}):
\begin{equation}\label{phi1}
    \phi - \phi_0 = \int^a_1 \frac{\sqrt{3B} ~da}{a(Aa^6+B)^{1/4}(3H^2_0\Omega_{m0}+(Aa^6+B)^{1/2})^{1/2}} \,,
\end{equation}
where $\Omega_{m0}\equiv \rho_{m0}/(3H_0^2)$ is the energy density parameter of matter and  we have set $a_0=1$. The
equation of state of Chaplygin gas is $ w \equiv p/\rho = -A/(A+ B/a^6)\ge-1$, then we get
\begin{equation}\label{AB1}
    B = -\frac{A(1+w_0)}{w_0} \,,
\end{equation}
and from the Friedmann equation (\ref{Fried equ}), we obtain
\begin{equation}\label{Fried1}
    3H^2_0 (1-\Omega_{m0}) = \sqrt{A+B} \,.
\end{equation}
From eq.(\ref{AB1}) and (\ref{Fried1}) we get
\begin{equation}\label{AB11}
    A = -w_0\bigg[3H_0^2(1-\Omega_{m0})\bigg]^2 \,, \quad B = (1+w_0)\bigg[3H_0^2(1-\Omega_{m0})\bigg]^2 \,,
\end{equation}
and then eq.(\ref{phi1}) becomes
\begin{eqnarray}
 \nonumber
  \phi - \phi_0 &=& \sqrt{3(1+w_0)}\int^a_1(1+w_0-w_0a^6)^{-1/4}\bigg(\frac{\Omega_{m0}}{1-\Omega_{m0}}
   + (1+w_0-w_0a^6)^{1/2}\bigg)^{-1/2}\frac{da}{a} \\
  \nonumber
   &=& \sqrt{\frac{\gamma_0}{3}} \int^{(\gamma_0-(\gamma_0-1)a^6)^{1/4}}_1 \frac{dy}{(y^2+r)^{1/2}}
   \left(\frac{1}{y^2-\sqrt{\gamma_0}} +\frac{1}{y^2+\sqrt{\gamma_0}}\right) \\
  \nonumber
  &&\\
   &=& \frac{\gamma_0^{1/4}}{\sqrt{3}}\left(\frac{\tan^{-1}{\left[\frac{y\sqrt{r-\gamma_0^{1/2}} }{\gamma_0^{1/4}
   \sqrt{r+y^2}}\right]}}{\sqrt{r-\gamma_0^{1/2}}}
   -\frac{\tanh^{-1}\left[\frac{y\sqrt{r+\gamma_0^{1/2}} }{\gamma_0^{1/4}
   \sqrt{r+y^2}}\right]}{\sqrt{r+\gamma_0^{1/2}}}\right)\Bigg|^{(\gamma_0-(\gamma_0-1)a^6)^{1/4}}_{y=1}
   \label{phi11} \,,
\end{eqnarray}
where $\gamma_0 = 1+w_0$, $r=\Omega_{m0}/(1-\Omega_{m0})<1$ and $y=(\gamma_0-(\gamma_0-1)a^6)^{1/4}$.  Therefore, if
$w_0 = -1$, $\phi$ is a constant during the evolution of the universe as the cosmological constant. When
\begin{equation}\label{cond1}
    a \gg \left[\frac{1+w_0-r^2}{w_0}\right]^{1/6} \,,
\end{equation}
we can neglect $r$ in eq.(\ref{phi1}) and get
\begin{equation}\label{phi11out}
   \phi - \phi_0 \approx -\frac{1}{\sqrt{3}}
   \tanh^{-1}\left(\frac{y^2}{\gamma_0^{1/2}}\right)\bigg|^{(\gamma_0-(\gamma_0-1)a^6)^{1/4}}_{y=1} \,.
\end{equation}

\subsection{Generalized Chaplygin gas model}
Although Chaplygin gas provides an interesting possibility for the unification of dark matter and dark energy. However,
it have to face some problems to explain some current observations such as it leads to the loss of power in CMB
anisotropies. This problem could be alleviated in the generalized Chaplygin gas model proposed in
ref.\cite{Bento:2002ps} (also see \cite{Bento:2002uh}) with equation of state
\begin{equation}\label{gen chaplygin gas}
    p = -\frac{A}{\rho^\sigma} \,,
\end{equation}
where $0 < \sigma\le 1$ and when $\sigma = 1$ it reduce to the pure Chaplygin gas (\ref{chaplygin gas}). Together with
energy conservation law $d(\rho a^3) = -pd(a^3)$, it gives
\begin{equation}\label{gen chaplygin gas energy density }
    \rho = \left(A + \frac{B}{a^{3(1+\sigma)}}\right)^{\frac{1}{1+\sigma}} \,,
\end{equation}
where $B$ is an integration constant. Hence, we can see that, for small and large $a$ it behaves like a dust and a
cosmological constant respectively, but in the intermediate phase, it looks like a mixture of a cosmological constant
with a "soft" matter whose equation of state is $p = \sigma\rho$ which can be obtained by expanding the pressure
(\ref{gen chaplygin gas}) and energy density (\ref{gen chaplygin gas energy density }) in subleading order:
\begin{equation}
    p \simeq  - A^{\frac{1}{1+\sigma}} + \frac{\sigma BA^{-\frac{\sigma}{1+\sigma}}}{1+\sigma}a^{-3(1+\sigma)}
    \,, \quad
    \rho \simeq A^{\frac{1}{1+\sigma}} + \frac{BA^{-\frac{\sigma}{1+\sigma}}}{1+\sigma}a^{-3(1+\sigma)}\,.
\end{equation}
One can also reconstruct a minimally coupled scalar field to mimic the behavior of generalized Chaplygin gas by
identify its energy density (\ref{gen chaplygin gas energy density }) and pressure (\ref{gen chaplygin gas}) to that of
the scalar field
\begin{eqnarray}
  \dot \phi^2 &=& \frac{B}{a^{3(1+\sigma)}\bigg(A+ \frac{B}{a^{3(1+\sigma)}}\bigg)^{\frac{\sigma}{1+\sigma}}} \label{phi rec 1}\,,\\
  V(\phi) &=&\frac{2A a^{3(1+\sigma)}+B}{2a^{3(1+\sigma)}\bigg(A+ \frac{B}{a^{3(1+\sigma)}}\bigg)^{\frac{\sigma}{1+\sigma}}}\,.
\end{eqnarray}
By using the Friedmann equation (\ref{Fried equ}), we  get the variation of $\phi$ in terms of the integration of $a$
from eq.(\ref{phi rec 1}):
\begin{equation}\label{phi2}
    \phi - \phi_0 = \int^a_1 \frac{\sqrt{3B} ~da}{a(Aa^{3(1+\sigma)}+B)^{\frac{\sigma}{2(1+\sigma)}}(3H^2_0\Omega_{m0}
    +(Aa^{3(1+\sigma)}+B)^{\frac{1}{1+\sigma}})^{1/2}} \,.
\end{equation}
The equation of state of generalized Chaplygin gas is $ w \equiv p/\rho = -A/(A+ B/a^{3(1+\sigma)})\ge-1$, then we get
\begin{equation}\label{AB2}
    B = -\frac{A(1+w_0)}{w_0} \,,
\end{equation}
which is the same as eq.(\ref{AB1}) and from the Friedmann equation (\ref{Fried equ}), we obtain
\begin{equation}\label{Fried2}
    3H^2_0 (1-\Omega_{m0}) = (A+B)^{\frac{1}{1+\sigma}} \,.
\end{equation}
From eq.(\ref{AB2}) and (\ref{Fried2}) we get
\begin{equation}\label{AB22}
    A = -w_0\bigg[3H_0^2(1-\Omega_{m0})\bigg]^{1+\sigma} \,, \quad B = (1+w_0)\bigg[3H_0^2(1-\Omega_{m0})\bigg]^{1+\sigma} \,,
\end{equation}
and then eq.(\ref{phi2}) becomes
\begin{eqnarray}
 \nonumber
  \phi - \phi_0 &=& \sqrt{3(1+w_0)}\int^a_1\bigg(1+w_0-w_0a^{3(1+\sigma)}\bigg)^{-\frac{\sigma}{2(1+\sigma)}}
  \bigg[r + \left(1+w_0-w_0a^{3(1+\sigma)}\right)^{\frac{1}{1+\sigma}}\bigg]^{-1/2}\frac{da}{a} \\
   &=& \sqrt{\frac{\gamma_0}{3}} \int^{\left[\gamma_0-(\gamma_0-1)a^{3(1+\sigma)}\right]^{\frac{1}{2(1+\sigma)}}}_1
   \frac{2y^{1+\sigma}dy}{(y^2+r)^{1/2}\left(y^{2(1+\sigma)}-\gamma_0\right)}
   \label{phi22} \,,
\end{eqnarray}
where $\gamma_0 = 1+w_0$, $r=\Omega_{m0}/(1-\Omega_{m0})$ and
$y=\left[\gamma_0-(\gamma_0-1)a^{3(1+\sigma)}\right]^{\frac{1}{2(1+\sigma)}}$. Therefore, eq.(\ref{phi22}) can be
analytically calculated when $r$ could be neglected, namely:
\begin{equation}\label{cond2}
    a \gg \left[\frac{\gamma_0-r^{1+\sigma}}{\gamma_0-1}\right]^{\frac{1}{3(1+\sigma)}} \,,
\end{equation}
and then we get
\begin{equation}\label{phi22out}
   \phi - \phi_0 \approx -\frac{2\sqrt{3}}{3(1+\sigma)}
   \tanh^{-1}\left(\frac{y^{1+\sigma}}{\gamma_0^{1/2}}\right)
   \bigg|^{\left(\gamma_0-(\gamma_0-1)a^{3(1+\sigma)}\right)^{\frac{1}{2(1+\sigma)}}}_{y=1} \,.
\end{equation}

\subsection{Modified generalized Chaplygin gas model}

Another candidate for the generalization of the Chaplygin gas called modified generalized Chaplygin gas or modified
Chaplygin gas model \cite{Benaoum:2002zs} is characterized by the following equation of state
\begin{equation}\label{mod gen chaplygin gas}
    p = A\rho-\frac{B}{\rho^\sigma} \,,
\end{equation}
where $A$, $B$ and $\sigma$ are constants and  $ 0 \le\sigma \le 1$. Thus, it looks like a mixture of two kinds of
fluids, one with equation of state $p=A\rho$ and the other one being the generalized Chaplygin gas. From eq.(\ref{mod
gen chaplygin gas}), one can see that it reduces to generalized Chaplygin gas when $A=0$ and to the perfect fluid if
$B=0$. Again, together with energy conservation law, it gives
\begin{equation}\label{mod gen chaplygin gas energy density }
    \rho = \left(\frac{B}{1+A} + \frac{C}{a^{3(1+A)(1+\sigma)}}\right)^{\frac{1}{1+\sigma}}
\end{equation}
where $C$ is an integration constant. Then, for small scale factor $a$, it behaves like a dust (if $A=0$) or radiation
(if $A=1/3$) with equation of sate $p=A\rho$ and energy density $\rho = C^{\frac{1}{1+\sigma}} a^{-3(1+A)}$, wile for
large $a$, it behaves like a cosmological constant. In the intermediate phase, it corresponds to the mixture of a
cosmological constant and a kind of fluid with equation of state $p = (\sigma + A + A\sigma )\rho$, which can be
obtained by expanding the pressure (\ref{mod gen chaplygin gas}) and energy density (\ref{mod gen chaplygin gas energy
density }) in subleading order:
\begin{eqnarray}
  p &\simeq&  - \left(\frac{B}{1+A}\right)^{\frac{1}{1+\sigma}}
              + \bigg(\sigma+A+A\sigma\bigg)\left(\frac{B}{1+A}\right)^{-\frac{\sigma}{1+\sigma}}\frac{C a^{-3(1+A)(1+\sigma)}}{1+\sigma}
    \,, \\
  \rho &\simeq& \left(\frac{B}{1+A}\right)^{\frac{1}{1+\sigma}}
              + \left(\frac{B}{1+A}\right)^{-\frac{\sigma}{1+\sigma}}\frac{C a^{-3(1+A)(1+\sigma)}}{1+\sigma}\,.
\end{eqnarray}
We also reconstruct a minimally coupled scalar field to mimic the behavior of generalized Chaplygin gas by identify its
energy density (\ref{mod gen chaplygin gas energy density }) and pressure (\ref{mod gen chaplygin gas}) to that of the
scalar field
\begin{eqnarray}
  \dot \phi^2 &=& \frac{(1+A)C}{a^{3(1+A)(1+\sigma)}\bigg(\frac{B}{1+A}+ \frac{C}{a^{3(1+A)(1+\sigma)}}\bigg)^{\frac{\sigma}{1+\sigma}}} \label{phi rec 3}\,,\\
  V(\phi) &=&\frac{2B(1+A)^{-1} a^{3(1+A)(1+\sigma)}+(1-A)C}{2a^{3(1+A)(1+\sigma)}\bigg(\frac{B}{1+A}+ \frac{C}{a^{3(1+A)(1+\sigma)}}\bigg)^{\frac{\sigma}{1+\sigma}}}\,.
\end{eqnarray}
By using the Friedmann equation (\ref{Fried equ}), we  get the variation of $\phi$ in terms of the integration of $a$
from eq.(\ref{phi rec 3}):
\begin{equation}\label{phi3}
    \phi - \phi_0 = \int^a_1 \frac{\sqrt{3(1+A)C} ~da}{a^{1+\frac{3}{2}A}\left(\frac{Ba^{3(1+A)(1+\sigma)}}{1+A}+C\right)^{\frac{\sigma}{2(1+\sigma)}}
    \left(3H^2_0\Omega_{m0} +\left(\frac{Ba^{3(1+A)(1+\sigma)}}{1+A}+C\right)^{\frac{1}{1+\sigma}}a^{-3A}\right)^{1/2}} \,.
\end{equation}
The equation of state of generalized Chaplygin gas is
\begin{equation}\label{eof3}
    w \equiv \frac{p}{\rho} = A - \frac{B(1+A)}{B+(1+A)Ca^{-3(1+A)(1+\sigma)}} \ge -1 \,,
\end{equation}
where we have used eq.(\ref{mod gen chaplygin gas}) and (\ref{mod gen chaplygin gas energy density }). From
eq.(\ref{eof3}), we get
\begin{equation}\label{AB3}
    C = \frac{B(1+w_0)}{(1+A)(A-w_0)} \,,
\end{equation}
and
\begin{equation}\label{AB3p}
    w_0' \equiv \frac{dw}{dz}\bigg|_{z=0} = \frac{3BC(1+A)^3(1+\sigma)}{\left[B+(1+A)C\right]^2} = 3(1+\sigma)(1+w_0)(A-w_0)
\end{equation}
where we have used eq.(\ref{AB3}) and $z = a^{-1}-1$ is the redshift. From the Friedmann equation (\ref{Fried equ}), we
obtain
\begin{equation}\label{Fried3}
    3H^2_0 (1-\Omega_{m0}) = \left(\frac{B}{1+A} + C\right)^{\frac{1}{1+\sigma}} \,.
\end{equation}
Using eq.(\ref{AB3}), (\ref{AB3p}) and (\ref{Fried3}), we get
\begin{equation}\label{AB33}
    B = (\tilde{A}-\gamma_0)\bigg[3H_0^2(1-\Omega_{m0})\bigg]^{1+\sigma} \,,
    \quad
    C = \frac{1+w_0}{\tilde{A}}\bigg[3H_0^2(1-\Omega_{m0})\bigg]^{1+\sigma} \,,
\end{equation}
where we have defined $\gamma_0 = 1+w_0$ and
\begin{equation}\label{AB33a}
   \tilde{A}\equiv 1+A = \frac{w_0'}{3(1+\sigma)\gamma_0} + \gamma_0 \,.
\end{equation}
Noticed that the case $A=0$ corresponds to $w'= 3\gamma_0(1-\gamma_0)(1+\sigma)$. Then eq.(\ref{phi3}) becomes
\begin{eqnarray}
 \nonumber
  \phi - \phi_0 &=& \sqrt{3\gamma_0}\int^a_1\bigg[\frac{\tilde{A}}{\gamma_0+ (\tilde{A}-\gamma_0)a^{3\tilde{A}(1+\sigma)}}\bigg]^{\frac{\sigma}{2(1+\sigma)}}
  \bigg[ra^{3(\tilde{A}-1)} + \left(\frac{\gamma_0+(\tilde{A}-\gamma_0)a^{3\tilde{A}(1+\sigma)}}{\tilde{A}}\right)^{\frac{1}{1+\sigma}}\bigg]^{-\frac{1}{2}}\frac{da}{a} \\
   &=& \sqrt{\frac{\gamma_0}{3}} \int^{\left[\frac{\gamma_0+(\tilde{A}-\gamma_0)a^{3\tilde{A}(1+\sigma)}}{\tilde{A}}\right]^{\frac{1}{2(1+\sigma)}}}_1
   \frac{2y^{1+\sigma}dy}{\left(y^2+ra^{3(\tilde{A}-1)}\right)^{1/2}\left(\tilde{A}y^{2(1+\sigma)}-\gamma_0\right)}
   \label{phi33} \,,
\end{eqnarray}
where $r=\Omega_{m0}/(1-\Omega_{m0})$ and
$y=\left[\frac{\gamma_0+(\tilde{A}-\gamma_0)a^{3\tilde{A}(1+\sigma)}}{\tilde{A}}\right]^{\frac{1}{2(1+\sigma)}}$.
Therefore, eq.(\ref{phi33}) can be analytically calculated when $r$ could be neglected, namely:
\begin{equation}\label{cond3}
    r \ll y^2 a^{-3(\tilde{A}-1)} \,, \quad \text{or} \quad
    a \gg \left[\frac{\gamma_0a^{3(\tilde{A}-1)(1+\sigma)}-\tilde{A}r^{1+\sigma}}{\gamma_0-\tilde{A}}\right]^{\frac{1}{3(1+\sigma)}}\,,
\end{equation}
and then we get
\begin{equation}\label{phi22out}
   \phi - \phi_0 \approx -\frac{2\sqrt{3}}{3(1+\sigma)\sqrt{\tilde{A}}}
   \tanh^{-1}\left(y^{1+\sigma}\sqrt{\frac{\tilde{A}}{\gamma_0}}\right)
   \Bigg|^{\left[\frac{\gamma_0+(\tilde{A}-\gamma_0)a^{3\tilde{A}(1+\sigma)}}{\tilde{A}}\right]^{\frac{1}{2(1+\sigma)}}}_{y=1} \,.
\end{equation}

\section{Varying alpha from Chaplygin gas models}
\begin{figure}[h]
\includegraphics[width=0.5\textwidth]{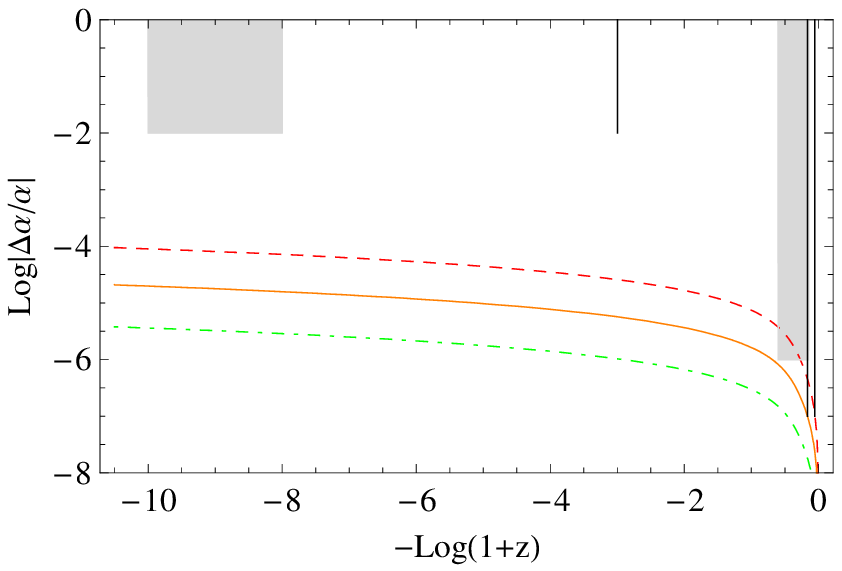}
\quad
\includegraphics[width=0.5\textwidth]{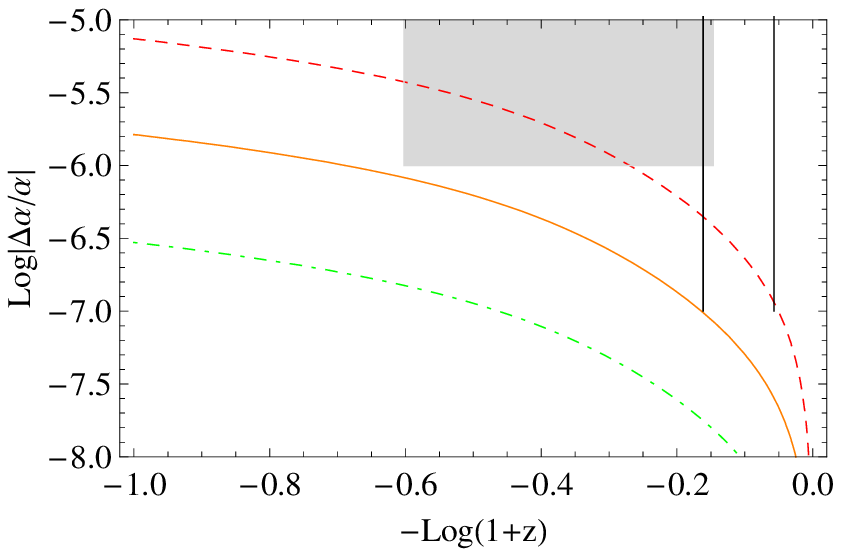}
\caption{\label{fig::case1}\emph{Chaplygin gas}: The variation of $\alpha$ during the evolution of the universe is
plotted, namely, $\log|\Delta\alpha/\alpha|$ vs. -$\log(1+z)$. Here we have used $w_0=0.99$, $\Omega_{m0}=0.27$ and the
solid, dashed and dot-dashed cures correspond to $\zeta = 1.1\times10^{-6}$, $\zeta =5.0\times10^{-6}$ and $\zeta
=0.2\times10^{-6}$ respectively. Only the curves that not overlaps the gray areas are phenomenologically viable.  }
\end{figure}

\begin{figure}[h]
\includegraphics[width=0.5\textwidth]{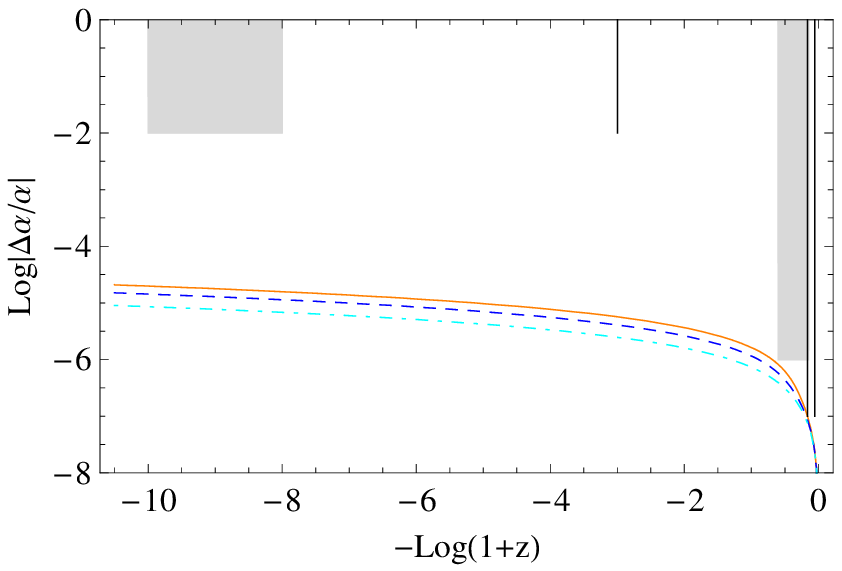}
\quad
\includegraphics[width=0.5\textwidth]{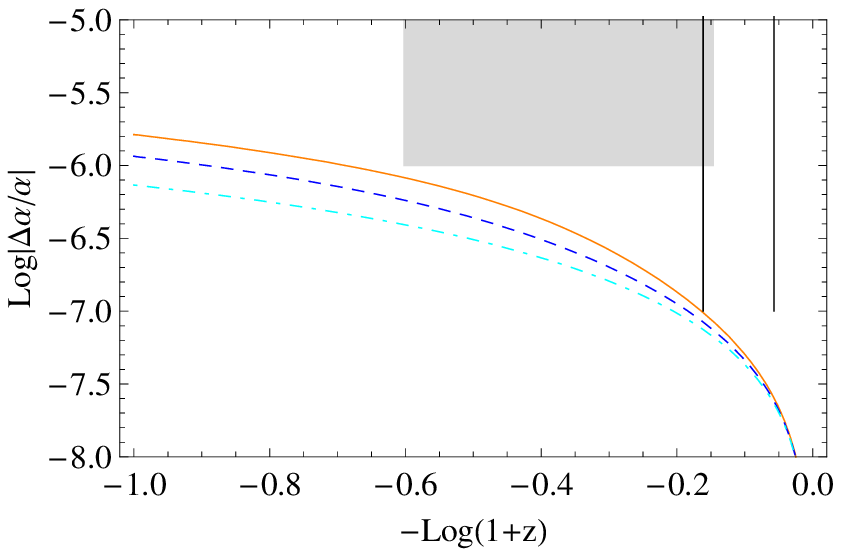}
\caption{\label{fig::case2}\emph{Generalized Chaplygin gas}: The variation of $\alpha$ during the evolution of the
universe is plotted, namely, $\log|\Delta\alpha/\alpha|$ vs. -$\log(1+z)$. Here we have used $w_0=0.99$,
$\Omega_{m0}=0.27$, $\zeta = 1.1\times10^{-6}$ and the solid, dashed and dot-dashed cures correspond to $\sigma =1.0$,
$\sigma=0.5$ and $\sigma=0.1$ respectively. Only the curves that not overlaps the gray areas are phenomenologically
viable.}
\end{figure}

\begin{figure}[h]
\includegraphics[width=0.5\textwidth]{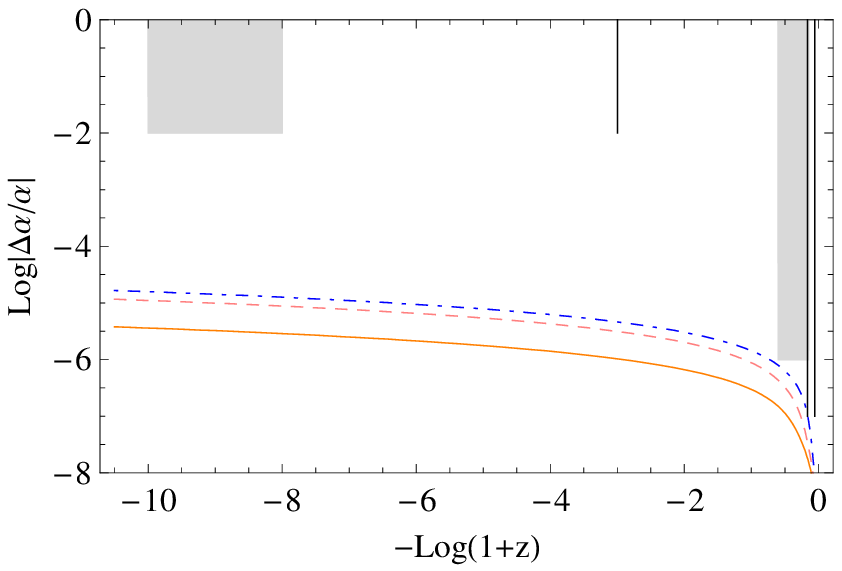}
\quad
\includegraphics[width=0.5\textwidth]{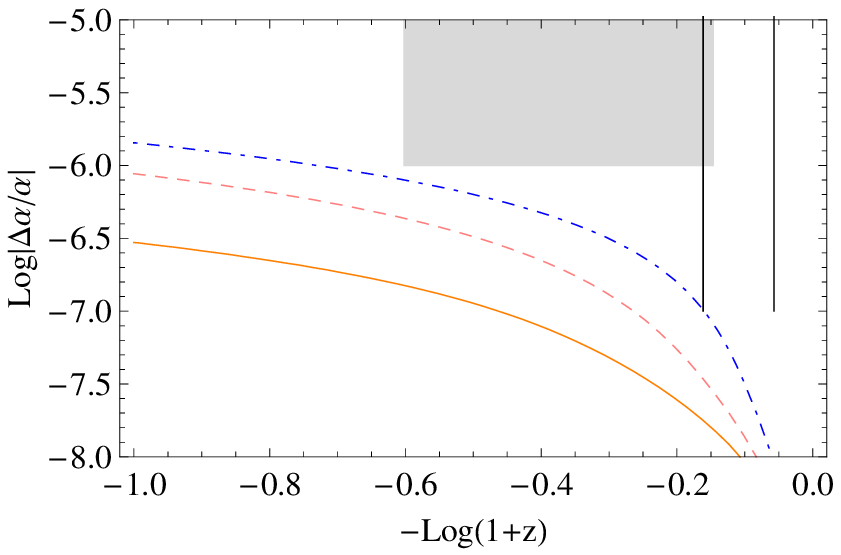}
\caption{\label{fig::case3}\emph{Modified generalized Chaplygin gas}: The variation of $\alpha$ during the evolution of
the universe is plotted, namely, $\log|\Delta\alpha/\alpha|$ vs. -$\log(1+z)$. Here we have used $w_0=0.99$,
$\Omega_{m0}=0.27$, $\zeta = 0.2\times10^{-6}$, $\sigma=1.0$ and the solid, dashed and dot-dashed cures correspond to
$A = 0$ (or $w'_0= 0.06$), $w'_0=0.12$ and $w'_0=0.24$ respectively. Only the curves that not overlaps the gray areas
are phenomenologically viable. }
\end{figure}

Although the form of the coupling between a scalar field and the electromagnetic field can be very complicated
\cite{Marra:2005yt}, in general, the observational results have indicated that the variation of $\alpha$ is small and
 $B_F$ can be approximated as a linear form in practice. Therefore, in this paper, we will take such a approximation
\begin{equation}\label{linear1}
    B_F(\phi) = 1-\zeta(\phi-\phi_0)\,,
\end{equation}
which corresponds to the choice of $\epsilon = \tau = 0$ and $q=1$ for the parameters in ref.\cite{Marra:2005yt}. From
the tests of the equivalence principle, the coupling is constrainted to be $|\zeta|<10^{-3}$. Then, the variation of
$\alpha$ is given by
\begin{equation}\label{variation of alpha}
    \bigg|\frac{\Delta\alpha}{\alpha}\bigg| = \bigg|\frac{1-B_F(\phi)}{B_F(\phi)}\bigg| \approx |\zeta(\phi-\phi_0)|.
\end{equation}
Here, we tried some different values of $\zeta$ to make all the constraints that mentioned in the introduction section
be satisfied. The variation of $\alpha$ is presented in Fig.\ref{fig::case1}, Fig.\ref{fig::case2} and
Fig.\ref{fig::case3} for the Chaplygin gas model, the generalized Chaplygin gas model and the modified generalized
Chaplygin gas model respectively.

From Fig.\ref{fig::case1}, one can see that all the constraints are respected for $\zeta\lesssim1.1\times10^{-6}$ in
the case of the Chaplygin gas model. For the case of generalized Chaplygin gas and a given value of $\zeta$, the
variation of $\alpha$ is getting smaller and smaller when the parameter $\sigma$ becomes small in the same situation.
In other words, the smaller $\sigma$ is, the larger upper bound of $\zeta$ is, see Fig.\ref{fig::case2}. Finally, in
the case of modified generalized Chaplygin gas, the variation of $\alpha$ becomes large when the present running of the
equation of state $w'_0$ is large. Thus, the upper bound of $\zeta$ should be smaller to satisfy constraints, see
Fig.\ref{fig::case3}.

\section{Conclusions}
In this paper, we have reconstructed the class of Chaplygin gas models to a kind of scalar field and study the
variation of the fine structure constant $\alpha$ driven by it. This phenomenon was found since ten years ago and
attracted many attentions. The resulting $|\Delta\alpha/\alpha|$ as a function of the redshif $z$ is presented in
Fig.\ref{fig::case1}, Fig.\ref{fig::case2} and Fig.\ref{fig::case3}. We only consider the case of linear coupling
between the scalar field and the electromagnetic field, because the variation of $\alpha$ is much small. The results
indicate that if the present observational value of the equation of state of the dark energy was not exactly equal to
$-1$, various parameters of the class of Chaplygin gas models are allowed to satisfy the observational constraints, as
well as the equivalence principle is also respected since it requires the constant $\zeta$ is much smaller than
$10^{-3}$ in all the case.

For the generalized Chaplygin gas, there is a parameter $0< \sigma \le 1$ in eq.(\ref{gen chaplygin gas}). We find that
when the smaller $\sigma$ is, the larger upper bound of $\zeta$ is. In the case of modified generalized Chaplygin gas,
the upper bound of $\zeta$ becomes more restricted when the running of equation of state $w'_0$ is large. It is worth
further studying, since the variation of fundamental constants during the cosmic time is a very interesting area.

\acknowledgments

This work is supported by National Science Foundation of China grant No.
10847153 and No. 10671128.

\end{document}